\documentclass[letter,traditabstract]{aa}  

\usepackage{graphicx}
\usepackage{txfonts}
\begin{document}

   \title{Very massive stars and nitrogen-emitting galaxies}

   \author{Jorick S. Vink \inst{1}}

   \institute{Armagh Observatory and Planetarium, College Hill,
              BT61 9DG Armagh, Northern Ireland\\
              \email{jorick.vink@armagh.ac.uk}
             }

   \date{Received 29 August 2023 / Accepted 16 October 2023}

 
  \abstract
{Recent studies of high-redshift galaxies with
James Webb Space Telescope (JWST), such as GN-z11 at $z=10.6,$  show unexpectedly significant amounts of nitrogen (N) in their spectra. As this phenomenology appears to extend to gravitionally lensed galaxies at Cosmic noon such as the Sunburst Arc at $z=2.37$, as well as globular clusters overall, we suggest that the common ingredient among them are very massive stars (VMSs) with zero-age main sequence (ZAMS) masses in the range of 100-1000\,$M_{\odot}$. 
The He {\sc ii} in the Sunburst Arc might also be the result of the disproportionally large contribution of VMS to the total stellar contribution. 
We analyse the pros and cons of the previous suggestions, including classical Wolf-Rayet (cWR) stars and supermassive stars (SMSs), to conclude that only our VMS alternative ticks all the relevant boxes. 
We discuss the VMS mass-loss history via their peculiar vertical evolution in the HR diagram resulting from a self-regulatory effect of these wind-dominated VMSs and we estimate that the large amounts of N present in star-forming galaxies may indeed result from VMSs. 
We conclude that VMSs should be included in population synthesis and chemical evolution models. Moreover, that it is critical for this to be done self-consistently, as a small error in their mass-loss rates would have dramatic consequences for their stellar evolution, as well as their ionising and chemical feedback.}


\maketitle
%

\section{Introduction}

High-redshift galaxies are now routinely studied with James Webb Space Telescope (JWST). One of the surprises gleaned from galaxies such as GN-z11 (Bunker et al. 2023) involves the discovery of large amounts of nitrogen (N), with elevated [N/O] ratios (Cameron et al 2023; Senchyna et al. 2023). A realistic option for this N-enhancement involves the creation and subsequent release of hydrogen (H) burning products via the CNO-cycle.

This would immediately hint at massive stars, such as classical helium (He) burning Wolf-Rayet stars (cWRs) to be the culprits, but in this phase N is actually being destroyed in the core. Instead, supermassive stars (SMSs with $M \simeq 10^4 M_{\odot}$; Denissenkov et al. 2014; Gieles et al. 2018) have been suggested (Charbonnel et al. 2023), as they may simultaneously be responsible for this N-enrichment in star-forming galaxies and the anti-correlations (Gratton et al. 2004; Bastian \& Lardo 2018) seen in globular clusters (GCs), as cWR winds are too fast to keep the enriched material in the GC potential well. The underlying reasoning to invoke a similar enrichment mechanism for GCs and N emitting galaxies, is that young massive clusters (YMCs) 
are thought to be the progenitors of GCs (Portegies-Zwart et al. 2010). 

However, thus far SMSs have mostly remained a hypothetical notion and when we observe YMCs in our Local Universe, such as R136 in the Large Magellanic Cloud (LMC), we do not observe SMSs, but what we do observe are very massive stars (VMSs) with masses over 100$M_{\odot}$. These VMSs are fundamentally different from normal massive O stars in several important respects, all related to the proximity to the Eddington ($\Gamma$) limit (Vink et al. 2015). The VMS winds are not only stronger, but when approaching the $\Gamma$-limit they become disproportionally high. Moreover, their winds are slower (see Sect.\,4) and, crucially, VMSs are close to chemical homogeneity (Gr\"afener et al. 2011; Yusof et al. 2013; Hirschi 2015). It is for these reasons that Vink (2018) originally suggested that VMSs are the most likely culprits responsible
for the pollution of young GCs, as here the chemical homogeneity is a natural consequence of their large convective cores and strong winds, and it does not rely on the highly uncertain efficiency of rotational mixing that is required for the fast-rotating massive star scenario (FRMS) of Decressin et al. (2007).

 We have recently implemented enhanced VMS winds in MESA stellar evolution models in the 100-1000 $M_{\odot}$ range (Sabhahit et al. 2022, 2023; Higgins et al. 2022, 2023) that are calibrated at the Vink \& Gr\"afener (2012) transition mass-loss rate, resulting in "vertical\footnote{Vertical evolution takes places when the luminosity drops quickly due to strong mass loss. Some earlier models (e.g. Yusof et al. 2013) showed vertical evolution for later evolution, when prematurely (when the surface $Y$ arbitrarily exceeds 0.4) switching to cWR mass-loss recipes (such as Nugis \& Lamers 2000) intended for more advanced evolutionary phases.
  Instead, the new Sabhahit et al. (2022, 2023) models employ enhanced winds due to multiple scattering already from the very start of H-burning, but only when winds are expected to be optically thick.}" rather than horizontal evolution in the HR diagram and strong mass evaporation effects. This leads to extreme enrichment of H-burning products of not only N, but also of key elements such as Na that take place during the CNO cycle of H burning (Higgins et al. 2023).

While spectral population synthesis models of extra-galactic populations, such as Starburst99 (Leitherer et al. 1999) and BPASS (Eldridge et al. 2017) and galactic chemical evolution (GCE) models (e.g. Kobayahsi \& Ferrera 2023) have been very insightful, they can only be as good as the quality of their stellar ingredients. For instance, Kobayashi \& Ferrara include rotating cWRs at low-$Z,$ whose existence is rather speculative. 
For many decades, one of the key missing ingredients  concerned VMSs that reveal themselves as luminous WNh stars.
In this letter, we argue that as VMSs dominate the He II line emission in local clusters like R136 (Crowther et al. 2016) they should also be considered as key sources of nitrogen (N) found in extreme galaxies at high-z and Cosmic Noon, as well as elements such as Sodium (Na) relevant for the observed Na-O anti-correlations in GCs (Vink 2018). We recap these arguments in Sect.\,3.

\section{VMS are WNh stars producing He {\sc ii} emission}

The VMSs are luminous ($\log (L/L_{\odot}) = 10^6 - 10^7$) H-burning WNh stars up to 200-300 $M_{\odot}$ (Crowther et al. 2010; Bestenlehner et al. 2014; Kalari et al. 2022) with very large mass-loss rates up to $10^{-4}$  $M_{\odot}$/yr (Crowther et al. 2010; Vink et al. 2011; Gr\"afener et al. 2011; Martins 2015). 

Vink et al. (2011) predicted not only enhanced mass-loss rates as a result of multiple scattering in optically thick winds (Vink \& Gr\"afener 2012), but also that He II emission lines\footnote{Note: the underlying physics of the optical 4686 \AA\ recombination line is the same of that of the UV 1640 \AA\ line (e.g. Gr\"afener \& Vink 2015).} would be expected to transform from an optically thin O-star absorption feature to an emission feature at the optically thin-thick mass-loss transition point. We note that while we are confident about our predicted mass-loss upturn for VMS, the absolute rates, as well as their wind velocities are still  uncertain (see the discussion in Vink et al. 2011). For this reason it was crucial that we were able to calibrate the absolute mass-loss rates at the $\eta = \tau = 1$ transition point\footnote{Here, $\eta$ refers to the efficiency of multi-line scattering and $\tau$ to the optical depth.} in Vink \& Gr\"afener (2012), which we have now implemented in stellar evolution models of Sabhahit et al. (2023).

While canonical O-type stars also produce nitrogen (N) in the CNO cycle of H-burning, expelling this material
in stellar winds, it is questionable canonical O-stars are the main polluter, as the mass-loss rates are modest, with $\dot{M}$ $\simeq$ $10^{-6}-10^{-7}$ $M_{\odot}$/yr. At older ages, some of the more massive O stars (and/or the more rapid rotators and/or those that lost H-rich material in a binary interaction) may become cWR stars. These 
stars can be N-rich or C-rich (and possibly O-rich) WN, WC, and WO subtypes (Crowther 2007).
These stars generally have little or no hydrogen left in their outer layers. 
By contrast VMSs are naturally close to chemically homogeneous and strongly N-enhanced and, as we argue in this letter, may be a key contributor to the enhanced N seen in the extreme N-emitting galaxies.

\begin{figure}
    \includegraphics[width=0.45\textwidth]{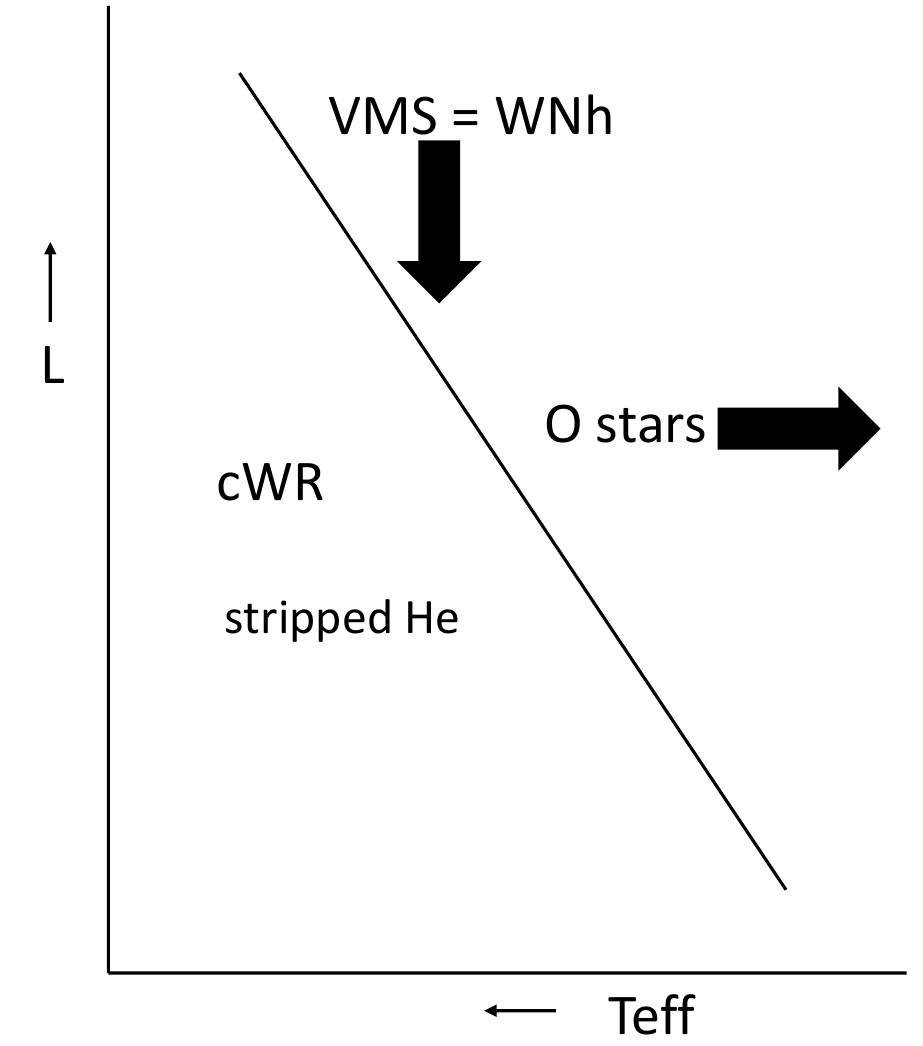}
    \caption{While canonical massive O-type stars are expected to evolve classically (i.e. horizontally) in the stellar HR diagram (HRD), VMS with $\Gamma$ enhanced winds are found to be evolving vertically downwards from the start of H burning as indicated by the downwards Thick arrow. They will likely produce classical WR (cWRs) that are on the hot side of the slanted line indicating the zero-age main sequence (ZAMS). At luminosities below a certain threshold (Shenar et al. 2020; Sander \& Vink 2020), stars may appear to be optically thin stripped He stars.}
    \label{fig:enter-label}
\end{figure}

While He II emission has been seen in star-bursting galaxies for decades (e.g. Brinchman et al. 2008), many in the extra-galactic community had traditionally divided He II emission into "stellar" WR emission (broad emission lines) and "nebular" (narrow emission) lines from photo-ionised gas (e.g. Cassata et al. 2013; Sobral et al. 2015; but see Gr\"afener \& Vink 2015). The He II radiation was often either attributed to hot pristine Pop III stars (e.g. Schaerer 2002) or in more recent times also to binary stripped He stars that could be present for a  longer time period (of order 50 Myrs; e.g. G\"otberg et al. 2020; Stanway et al. 2016). The issue with the majority of He stars being stripped in a binary is that these hot He products have not actually been observed in sufficiently large numbers. 

While such a galaxy as the LMC contains hundreds of classical WR stars above a given luminosity limit (Shenar et al. 2020), due to the shape of the initial mass function (IMF) many more lower-mass stripped He stars should be present, if the main formation channel for hot helium stars was due to binary stripping. Instead, only tens have been discovered (Drout et al. 2023). 

It could be that binary stripping is still relevant in a strict stellar evolutionary sense, involving partial stripping of the H envelope (e.g. Klencki et al. 2022, Gilkis et al. 2019), but in that case, the stripped He stars would be much cooler than the cWR stars and partially stripped He stars would hardly contribute significantly to the He II ionising radiation budget of galaxies.

\section{Alternatives}

One of the key reasons VMSs are a prime candidate is that they are known to exist, while some of the alternatives (e.g. SMSs) might not. 
Let us recap some of the pros and cons of the various scenarios for the proposed N (and Na) pollution in GCs, starting with the AGB scenario. Their main advantage over massive stars is that their winds are intrinsically slow (e.g. d'Ercole et al. 2008). However, 
their nucleo-synthetic evidence does not seem to be optimal, in particular as AGB stars give an Na-O correlation instead of the observed anti-correlation (e.g. Denissenkov et al. 2014). 
Massive stars are more in line with what is observed in terms of the Na-O anti-correlation, but they generally exhibit fast radiative winds. For these reasons, adaptations have been proposed to relieve this issue. 

One suggestion was to consider rapidly rotating stars, leading to slower winds at the equator (Decressin et al. 2007), and the other concerned binary interaction (de Mink et al. 2009). Issues with the former would be that it relies on the efficiency of rotationally induced chemical homogeneity (Vink \& Harries 2017\footnote{If rotating cWR stars would undergo rotationally induced chemical homogeneous evolution (CHE) one should expect to find a higher incidence of rotating cWR stars in the low $Z$ environment of the SMC than in higher $Z$ environments of the Galaxy and the LMC. However, the incidence of line polarisation effects due to rotation were not found to be any higher in the SMC than at higher $Z$, challenging the physics underlying rotationally-induced CHE (Vink \& Harries 2017.}), 
while an issue of the interacting binary model would be that the mass loss occurs at a fine-tuned moment (Bastian \& Lardo 2018). 

Regarding N-emitting galaxies, massive stars have been revived in the shape of cWR stars (see also Isobe et al. 2023; Marques-Chaves et al. 2023; Maiolino et al. 2023), partly because massive star yields are so attractive, and He {\sc ii} emission has been seen in extreme galaxies (e.g. Senchyna et al. 2023). 
However, the fundamental problem with employing cWR stars is that during core He-burning, nitrogen is actually being destroyed at the expense of C and O so that when layers that have been processed by He-burning reactions appear at the surface, the material injected into the interstellar medium is N-depleted, making it unlikely that cWRs are the culprit. 
Charbonnel et al. (2023) came to similar conclusions, although their preferred nitrogen sources were the SMSs of $\simeq $ $10^4$ $M_{\odot}$, which  have yet to be directly discovered in nature.

\section{Key points supporting VMSs}

Firstly, VMSs have huge convective cores leading to chemical homogeneity. This implies that, independently of their assumed rotations rates, they mix their interior H-burning products (e.g.
N and Na) extremely efficiently.
A second strength of the VMS scenario for explaining the extreme N enhancement is that VMSs are expected to start their H-burning evolution with very large L/M ratios (and Eddington parameter $\Gamma_{\rm e}$ that scale with $L/M$). 

\begin{equation}
    \Gamma_{\rm e} = \frac{\kappa_{\rm e} L}{4 \pi c G M}
.\end{equation}

These VMSs are subjected to huge amounts of mass loss already from the very start of core H-burning, which is naturally accompanied by reduced stellar wind velocities ($v_{\infty}$) due to this same proximity to the Eddington limit (Castor et al. 1975; Puls et al. 2008), as per:\ 
\begin{equation}
    v_{\infty} \propto v_{\rm esc} \propto \sqrt{\frac{2 G M (1\,-\,\Gamma_{\rm e})}{R}} 
,\end{equation}
where $v_{\rm esc}$ is the effective escape speed.
As VMSs naturally evaporate (Vink 2018; Higgins et al. 2022) and evolve vertically downwards in the HRD towards lower luminosity (Sabhahit et al. 2022), they get further away from the Eddington limit for radiation pressure. 

As VMS evolution models show a very rapid increase in the amounts of surface N by a factor 10; see Fig.\,2) material, prior to the release of He by smaller but significant amounts (see Fig.\,2, by a factor $<2$ and in line with GC observations). This means that this early wind N (as well as Na) production could take place at elevated rates in comparison to He enhancement when the mass-loss rates has dropped by a factor 3-5. 
This circumvents one of the key problems of alternative pollutor models for the He enhancement in GCs, as there would probably need to be some moderate He enrichment to explain the HRD locations of multiple main sequences (Piotto et al. 2005), as well as the distribution of stars along the horizonal branch (Norris et al. 1981, Chantereau et al. 2016). However, this amount of He should not be as extremely large as that produced in other models of massive stars, such as the fast-rotating massive star model of Decressin et al. (2007). 

\begin{figure}
    \includegraphics[width=0.45\textwidth]{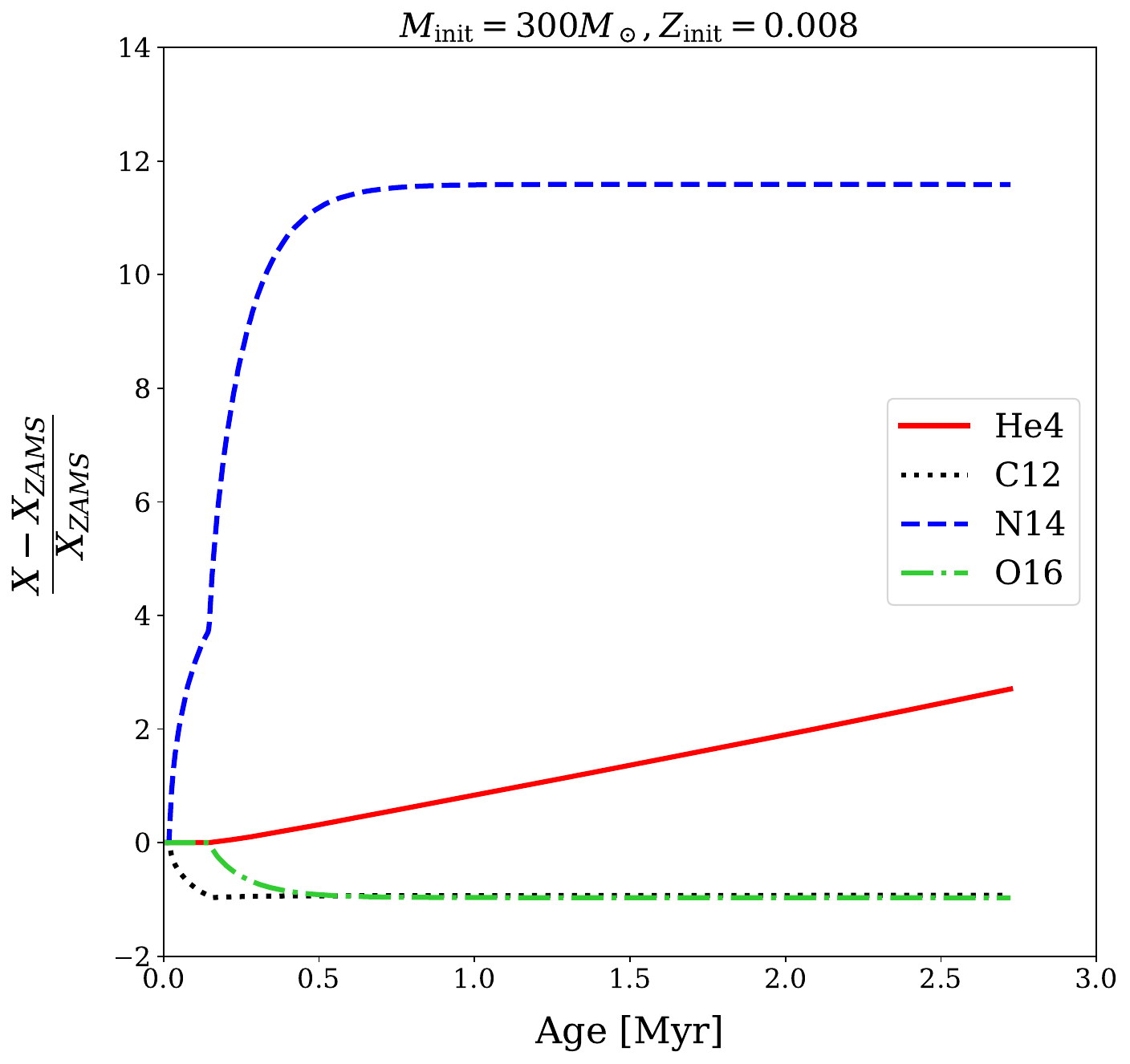}
    \caption{Relative surface enrichment for a 300\,$M_{\odot}$ VMSs at sub-solar LMC-like metallicity. The VMS MESA stellar evolution models are from Sabhahit et al. (2023).}
    \label{fig:abund}
\end{figure}

The key point is that helium production takes place when (i) mass-loss rates have already been reduced as a natural result of dropping luminosities and (ii) wind velocities are expected to be higher. The reason for these naturally increased wind velocities is two-fold. First, in hydrodynamically consistent models, lower mass-loss rates imply lower wind densities at the sonic point, leading  (in the case of optically thick spectral lines) to higher wind velocities (Vink 2022). Secondly, as the stars have naturally evolved away from their Eddington limit, the terminal wind velocity is yet higher again (see Eq.\,2).

To summarise, He production takes place at reduced levels as well as larger outflow speeds than the earlier production of species such as N and Na, naturally leading to a relatively higher N and Na production than He production. In addition, VMS wind outflow speeds are expected to be lower at lower $Z$ (Gr\"afener \& Vink 2015) and lower effective temperature (Vink 2018).

The exact evolutionary and population scenarios still need to be worked out, but this would only make sense when VMS evolution models employ the correct wind physics. However, such methods are still in development (Vink 2022; Moens et al. 2022; Sabhahit et al. 2023; Sander et al. 2023).

\section{Quantification}

While the final numbers of N enrichment will have to await more appropriate stellar evolution models that include the VMS mass-loss self-consistently, we did make some order of magnitude estimates, as described below.

Mestric et al. (2023) recently performed population synthesis of the Sunburst Arc, a gravitationally lensed source at Cosmic Noon (z=2.37). Considering the He {\sc ii} emission line to be due to VMSs (Vink et al. 2011), these authors combined new stellar atmosphere and evolution models of Martis \& Palacios (2022), arriving at a total number of VMSs of 400. Considering that the Sunburst Arc has a higher cluster mass than R136 in the LMC by approximately 100 times ($10^7$ $M_{\odot}$ versus $10^5$ $M_{\odot}$), the number is in broad agreement with the spatially resolved number of seven VMSs in R136 (Crowther et al. 2016). Pascale et al. (2023) derived 500\,$M_{\odot}$ of nitrogen for the Sunburst Arc. 
We consider whether VMSs could indeed produce such a large amount of nitrogen.

To address this question, we first analysed the surface N abundances of VMSs from our own evolutionary models, while including wind enhancement computed with MESA (Sabhahit et al. 2022). These LMC models show that VMSs in the range 100-500 $M_{\odot}$ are of the order  of 1\% of nitrogen (by mass) at their surface. Given that {LMC
VMSs of 300 $M_{\odot}$ have mass-loss rates of $10^{-4}$\,$M_{\odot}/$yr (Crowther et al. 2010; Vink et al. 2011) from the start of H burning, {these Sabhahit et al. LMC models} produce $\sim$1\,$M_{\odot}$ of nitrogen on the evolutionary H-burning timescale of $\sim$2 mega years and prior to there being any notable He enrichment. In other words, VMSs may indeed produce the large amount of nitrogen in the Sunburst Arc at Cosmic noon and they may also be the main source of N emission in galaxies such as GN-z11 at high redshift.

\section{Role of metallicity $Z$}

Nitrogen enrichment is expected to scale with the wind strength and therefore the initial
iron (Fe) metallicity, $Z$ (Vink \& de Koter 2005; Eldridge \& Vink 2006).
This would naively make it harder for the winds from VMSs to play this role for metallicities that are too low. 
Additionally, there is a maximum possible surface N-enrichment resulting from layers appearing at the surface having undergone CNO processing.
For the Sunburst Arc (discussed in the previous section) the best estimates are LMC-like (i.e. 0.4\,$Z/Z_{\odot}$) Mainali et al. 2022; Chisholm et al. 2019). For this reason, the proposed mechanism is most convincing here. For the high-redshift galaxy GN-z11,  numbers of order of 0.1 $Z_{\odot}$ have been discussed (Cameron et al. 2023) and this is still within the high $Z$ regime of Sabhahit et al. (2023), where the $\Gamma$ dependence dominates over the $Z$ dependence.
For lower GC-like metallicities, the task at hand for VMSs to be the main culprit will be harder -- but not impossible. 

First, it should be noted that alternative wind sources, such as rotating cWR stars and SMSs would face the exact same $Z$ dependence challenge as our VMSs suggested here. Although this challenge is less severe for VMSs than for cWRs, as VMSs are close to the Eddington limit, where the $\Gamma$ dependence dominates over the $Z$-dependence. 

There is a second relevant issue in that star formation at the top end of the mass function itself is also $Z$-dependent. 
Arguments for a more top-heavy initial mass function (IMF) due to less cooling in the Early Universe are decades old (Bromm et al. 1999; Abel et al. 2000), but there are newer arguments on the grounds of wind physics: Vink (2018) argued that weaker radiation-driven winds at lower $Z$ could naturally result in a higher upper mass limit at a lower $Z$.

For these various reasons it is not yet possible to predict how the overall N enhancement of VMSs would change with $Z$. It is, for instance, not inconceivable that the increasing upper mass limit from Vink (2018) at lower $Z$ would extend into the regime of SMSs (Woods et al. 2021, Haemmerl\'e 2021).

\section{Summary and discussion}

In this letter, we suggest that the most relevant stellar source of N enrichment is the same one as that responsible for He {\sc ii} emission, namely, the strong winds of VMSs. These stars have the advantage over SMSs in that they are known to exist in nature, and with respect to classical WR stars that they produce the correct nucleo-synthesis, namely, strong N enhancement, which would be destroyed during the He-burning of classical WR stars.

The key point is that VMSs may elegantly link together the enrichment of N in Globular clusters (Vink 2018; Higgins et al. 2023) and high-redshift galaxies. Key sources that link these objects are gravitionally lensed galaxies such as the Sunburst Arc at Cosmic Noon, which show both the N-enrichment (Pascale et al. 2023) and the He {\sc ii} emission that are the likely sign of VMSs (Mestric et al. 2023). 

Future population synthesis efforts based on high-redshift galaxies should not only focus on simply including VMSs, which is only beginning  (Wofford et al. 2023; Smith et al. 2023), but by doing so self-consistently. While stellar evolution models for VMSs exist for both high- and low-$Z$ (Belkus et al. 2007, Yungelson et al. 2008, Yusof et al. 2013, K\"ohler et al. 2015, Chen et al. 2015, Szecsi et al. 2015; Martinet et al. 2023), they generally do not yet include the appropriate optically thin-to-thick mass-loss physics. On a more positive note, a new $L/M$-dependent mass-loss framework for our understanding of VMSs is under construction (Sabhahit et al. 2023).

\begin{acknowledgements}
I would like to thank the anonymous referee for constructive comments and questions that helped to improve the paper. I would like to extend my gratitude to my students and colleagues in the Armagh Mdot-group, in particular Gautham Sabhahit and Erin Higgins for their help and comments on an earlier draft. 
\end{acknowledgements}

\end{document}